# Series solution of a ten-parameter second order differential equation with three regular and one irregular singularities

A. D. Alhaidari

*Saudi Center for Theoretical Physics, P. O. Box 32741, Jeddah 21438, Saudi Arabia*

**Abstract**: We introduce a ten-parameter ordinary linear differential equation of the second order with four singular points. Three of these are finite and regular whereas the fourth is irregular at infinity. We use the tridiagonal representation approach to obtain a solution of the equation as bounded infinite series of square integrable functions that are written in terms of the Jacobi polynomials. The expansion coefficients of the series satisfy three-term recursion relation, which is solved in terms of a modified version of the continuous Hahn orthogonal polynomial.

## I. Introduction

Recently, we introduced the following hypergeometric-type and confluent-hypergeometric-type differential equations

$$\left\{ x(1-x)\frac{d^2}{dx^2} + \left[a - x(a+b)\right]\frac{d}{dx} + \frac{A_+}{x} + \frac{A_-}{1-x} + A_1 x \right\} y(x) = A_0 y(x), \quad (1a)$$

$$\left[ x\frac{d^2}{dx^2} + (a+bx)\frac{d}{dx} + \frac{A_+}{x} + A_1 x \right] y(x) = A_0 y(x), \quad (1b)$$

where $\{a, b, A_0, A_1, A_\pm\}$ are real parameters. The original equations are obtained by taking $A_\pm = A_1 = 0$ with $A_0$ a discrete parameter (quadratic and linear in an integer $n$, respectively). We obtained solutions of these two equations as infinite or finite series of square integrable functions with the expansion coefficients being orthogonal polynomials of continuous and/or discrete spectra [1]. Most of those are well-known polynomials while few are not. We used the Tridiagonal Representation Approach (TRA) [2], which was inspired by the *J*-matrix method, in deriving these solutions. The *J*-matrix method was proposed in the mid 1970's by Yamani *et al* to handle the scattering problem in quantum mechanics [3-6]. Originally, the method was considered a purely physical technique but then Ismail and his collaborators turned it into a mathematical technique (see, for example, recent studies in Refs. [7-10]). More recently, we used the TRA to find solutions of the following nine-parameter Heun-type differential equation [11]

$$x(x-1)(x-r)\frac{d^2 y(x)}{dx^2} + \left\{ x^2(a+b+c) - x\left[a + c + r(a+b)\right] + ra \right\}\frac{dy(x)}{dx}$$
$$+ \left[ \frac{A}{x} + \frac{B}{x-1} + \frac{C}{x-r} + xD - E \right] y(x) = 0 \quad (2)$$

where $\{r,a,b,c,A,B,C,D,E\}$ are real parameters with $r \neq 0,1$. The original Heun equation corresponds to $A=B=C=0$ and $D=\alpha\beta$ with the regularity condition at infinity that $\alpha+\beta+1=a+b+c$ [12,13]. We showed that if the differential equation parameters $A$ and $B$ are below the critical values, $4A \leq r(1-a)^2$ and $4B \leq (1-r)(1-b)^2$, then we could find three classes of series solutions of Eq. (2) which are written in terms of orthogonal polynomials.

In the this paper, we try to generalize our work in [11] by finding series solutions for the following 2nd order linear ordinary differential equation

$$\left\{\pi_3(x)\frac{d^2}{dx^2}+[\pi_2(x)+\pi_0\pi_3(x)]\frac{d}{dx}+\frac{\hat{\pi}_2(x)}{\pi_3(x)}+\pi_1(x)\right\}y(x)=0, \tag{3}$$

where $\pi_n(x)$ and $\hat{\pi}_n(x)$ are polynomials in $x$ of degree $n$. We take $x \geq 0$ and assume that $\pi_3(x)$ has distinct real roots $\{\alpha,\beta,\gamma\}$ ordered as $\alpha<\beta<\gamma$. Thus, we can write $\pi_3(x)=(x-\alpha)(x-\beta)(x-\gamma)$. Now, making the linear displacement $x \to x+\alpha$ in Eq. (3) then dividing the equation by $\beta-\alpha$, we obtain

$$\left\{x(1-x)(r-x)\frac{d^2}{dx^2}+[p_2(x)+x(1-x)(r-x)\pi_0]\frac{d}{dx}+\frac{\hat{p}_2(x)}{x(1-x)(r-x)}+p_1(x)\right\}y(x+\alpha)=0, \tag{4}$$

where $r=\frac{\gamma-\alpha}{\beta-\alpha}$, which is larger than one. Moreover, we have defined the new polynomials $p_1(x)=\frac{\pi_1(x+\alpha)}{\beta-\alpha}$, $p_2(x)=\frac{\pi_2(x+\alpha)}{\beta-\alpha}$ and $\hat{p}_2(x)=\frac{\hat{\pi}_2(x+\alpha)}{(\beta-\alpha)^2}$. Now, we write these polynomials as

$$p_2(x)=\tilde{a}+\tilde{b}x+\tilde{c}x^2=x(1-x)(r-x)\left[\frac{a}{x}-\frac{b}{1-x}-\frac{c}{r-x}\right], \tag{5a}$$

$$\hat{p}_2(x)=\tilde{A}+\tilde{B}x+\tilde{C}x^2=x(1-x)(r-x)\left[\frac{A}{x}-\frac{B}{1-x}-\frac{C}{r-x}\right], \tag{5b}$$

$$p_1(x)=xD-E. \tag{5c}$$

where $\{a,b,c,A,B,C,D,E\}$ are real parameters, $a=r^{-1}\tilde{a}$, $b=\frac{\tilde{a}+\tilde{b}+\tilde{c}}{1-r}$ and $c=\frac{\tilde{a}+r\tilde{b}+r^2\tilde{c}}{r(r-1)}$. Analogous relations hold among the parameters of the polynomial $\hat{p}_2(x)$. Therefore, we can rewrite Eq. (4) as the following ten-parameter equation

$$\left\{x(1-x)(r-x)\left[\frac{d^2}{dx^2}+\left(\frac{a}{x}-\frac{b}{1-x}-\frac{c}{r-x}+d\right)\frac{d}{dx}\right]+\frac{A}{x}-\frac{B}{1-x}-\frac{C}{r-x}+xD-E\right\}\hat{y}(x)=0. \tag{6}$$

where $d=\pi_0$ and $\hat{y}(x)=y(x+\alpha)$. This differential equation has four singularities at $x=\{0,1,r,\infty\}$. If $d=0$, then these are all regular singular points and Eq. (6) becomes identical to the Heun-type equation (2). However, if $d \neq 0$ then the singular point at infinity becomes irregular (essential singularity). Therefore, we limit our study here to the case $d \neq 0$ and look for solutions of equation (6) that could be written as series of square integrable functions and as follows

$$\hat{y}(x) = \sum_{n=0}^{\infty} f_n \phi_n(x), \tag{7}$$

where $\{f_n\}$ is a properly chosen set of expansion coefficients. Due to the essential singularity at infinity and regularity of the singular points at $x = \{0, 1, r\}$, we propose the following square integrable basis functions

$$\phi_n(x) = g_n x^\alpha (1-x)^\beta (r-x)^\gamma e^{-\delta x} P_n^{(\mu,\nu)}(x), \tag{8}$$

where $g_n$ is a normalization constant and $P_n^{(\mu,\nu)}(x)$ is the Jacobi polynomial which is defined here as

$$P_n^{(\mu,\nu)}(x) = \frac{\Gamma(n+\mu+1)}{\Gamma(n+1)\Gamma(\mu+1)} {}_2F_1\left(-n, n+\mu+\nu+1; \mu+1; 1-x\right), \tag{9}$$

with $\mu > -1$, $\nu > -1$ and $x \in [0, +1]$. This definition is obtained by the replacement $x \to 2x - 1$ in the classical definition. The six newly defined basis parameters $\{\alpha, \beta, \gamma, \delta, \mu, \nu\}$ in (8) will be related to the differential equation parameters by square integrability of $\{\phi_n(x)\}$ and by TRA requirements on the expansion coefficients $\{f_n\}$ that will become clear shortly below. Square integrability is with respect to some positive measure, $d\zeta(x)$. That is, $\langle \phi_n | \phi_m \rangle \equiv \int \phi_n(x) \phi_m(x) d\zeta(x) = \Omega_{nm}$, which is finite. In the following section, we show how to use the TRA to obtain a series solution of the differential equation (4).

## II. TRA Solution

The differential equation and differential property of the Jacobi polynomials together with their recursion relation make it possible for the action of the second order differential operator (6) on the basis element $\phi_n(x)$ to produce terms proportional to $\phi_n(x)$ and $\phi_{n\pm 1}(x)$ with constant multiplicative factors. However, for that to happen, certain relationships among the parameters of the bases and those of the differential equation and constraints thereof must be satisfied. In that case, we can write $\mathcal{J}\phi_n(x) = \eta(x)[a_n \phi_n(x) + t_n \phi_{n+1}(x) + s_n \phi_{n-1}(x)]$ where Eq. (6) is written as $\mathcal{J} \hat{y}(x) = 0$ and $\eta(x)$ is an entire function. Therefore, substituting $\hat{y}(x)$ as given by Eq. (7) in the differential equation gives

$$\eta(x) \sum_{n=0}^{\infty} f_n [a_n \phi_n(x) + t_n \phi_{n+1}(x) + s_n \phi_{n-1}(x)] = 0. \tag{10}$$

Consequently, the expansion coefficients $\{f_n\}$ will satisfy the following three-term recursion relation

$$u_n z f_n(z) = w_n f_n(z) + t_{n-1} f_{n-1}(z) + s_{n+1} f_{n+1}(z), \tag{11}$$

for $n = 1, 2, 3, ...$ and where we wrote $a_n := w_n - zu_n$ with $z$ being some proper function of the differential equation parameters. This allows us to take full advantage of the many powerful techniques and analytic properties of orthogonal polynomials in the solution [14-17]. If the basis elements are orthonormal (i.e., $\langle \phi_n | \phi_m \rangle = \delta_{nm}$) then $z = E$ and $u_n = 1$ but this is not always the case. The recursion coefficients $\{u_n, w_n, s_n, t_n\}$ depend on the differential equation parameters and on $n$ but are independent of $z$ and such that $t_n s_n > 0$ for all $n$. Therefore, the solution $f_n(z)$ of (11) becomes a polynomial of degree $n$ in $z$ modulo an overall factor that depends on $z$ but is independent of $n$. That is, if we write $f_n(z) = f_0(z) P_n(z)$ then $P_0(z) = 1$. Moreover, defining $t_{-1} \equiv 0$, then Eq. (11) for $n = 0$ makes $P_1(z) = (zu_0 - w_0)/s_1$ and the recursion relation gives the polynomials $\{P_n(z)\}$ explicitly for all degrees starting with the initial values $P_0(z)$ and $P_1(z)$. Furthermore, the positive definite weight function for these polynomials becomes $[f_0(z)]^2$ [2,18]. Thus, the solution of the differential equation (3) is equivalent to the solution of the three-term recursion relation (11). In the following, we show how to derive this three-term recursion relation.

If we write equation (6) as $\mathcal{J}\hat{y}(x) = 0$, then the action of the 2nd order differential operator $\mathcal{J}$ on the basis element $\phi_n(x)$ as given by (8) becomes

$$\mathcal{J}\phi_n(x) = g_n x^{\alpha+1}(1-x)^{\beta+1}(r-x)^{\gamma+1} e^{-\delta x} \left[ \left( \frac{d}{dx} + \frac{\alpha}{x} - \frac{\beta}{1-x} - \frac{\gamma}{r-x} - \delta \right)^2 + \left( \frac{a}{x} - \frac{b}{1-x} - \frac{c}{r-x} + d \right) \times \right. \quad (12)$$
$$\left. \left( \frac{d}{dx} + \frac{\alpha}{x} - \frac{\beta}{1-x} - \frac{\gamma}{r-x} - \delta \right) + \frac{1}{x(1-x)(r-x)} \left( \frac{A}{x} - \frac{B}{1-x} - \frac{C}{r-x} + xD - E \right) \right] P_n^{(\mu,\nu)}(x)$$

Expanding and collecting similar terms give

$$\mathcal{J}\phi_n(x) = g_n x^{\alpha+1}(1-x)^{\beta+1}(r-x)^{\gamma+1} e^{-\delta x} \left[ \frac{d^2}{dx^2} + \left( \frac{2\alpha+a}{x} - \frac{2\beta+b}{1-x} - \frac{2\gamma+c}{r-x} + d - 2\delta \right) \frac{d}{dx} \right.$$
$$+ \delta(\delta - d) + (d - 2\delta)\left( \frac{\alpha}{x} - \frac{\beta}{1-x} - \frac{\gamma}{r-x} \right) - \delta\left( \frac{a}{x} - \frac{b}{1-x} - \frac{c}{r-x} \right) + \frac{\alpha(\alpha-1+a)}{x^2}$$
$$+ \frac{\beta(\beta-1+b)}{(1-x)^2} + \frac{\gamma(\gamma-1+c)}{(r-x)^2} - \frac{2\alpha\beta + \alpha b + \beta a}{x(1-x)} - \frac{2\alpha\gamma + \alpha c + \gamma a}{x(r-x)} + \frac{2\beta\gamma + \beta c + \gamma b}{(1-x)(r-x)} \quad (13)$$
$$\left. + \frac{1}{x(1-x)(r-x)} \left( \frac{A}{x} - \frac{B}{1-x} - \frac{C}{r-x} + xD - E \right) \right] P_n^{(\mu,\nu)}(x)$$

Using the differential equation of the Jacobi polynomial,

$$\left\{ x(1-x)\frac{d^2}{dx^2} + \left[ \nu + 1 - x(\mu + \nu + 2) \right] \frac{d}{dx} + n(n + \mu + \nu + 1) \right\} P_n^{(\mu,\nu)}(x) = 0, \quad (14)$$

turns Eq. (13) into the following

–4–

$$\mathcal{J}\phi_n(x) = g_n x^\alpha (1-x)^\beta (r-x)^{\gamma+1} e^{-\delta x} \left\{ \left( \frac{2\alpha + a - \nu - 1}{x} - \frac{2\beta + b - \mu - 1}{1-x} - \frac{2\gamma + c}{r-x} + d - 2\delta \right) x(1-x) \frac{d}{dx} \right.$$

$$-n(n+\mu+\nu+1) - (2\alpha\beta + \alpha b + \beta a) + \delta(\delta - d)x(1-x) + (d-2\delta)\left[ \alpha - \gamma(r-1) - (\alpha+\beta+\gamma)x + \frac{r(r-1)\gamma}{r-x} \right]$$

$$-\delta\left[ a - c(r-1) - (a+b+c)x + \frac{r(r-1)c}{r-x} \right] + \alpha(\alpha - 1 + a)\left( \frac{1}{x} - 1 \right) + \beta(\beta - 1 + b)\left( \frac{1}{1-x} - 1 \right) \quad (15)$$

$$-(2\alpha\gamma + \alpha c + \gamma a)\left( 1 - \frac{r-1}{r-x} \right) + (2\beta\gamma + \beta c + \gamma b)\left( \frac{r}{r-x} - 1 \right) + \gamma(\gamma - 1 + c)\left[ -1 + \frac{2r-1}{x-r} - \frac{r(r-1)}{(r-x)^2} \right]$$

$$\left. + \frac{1}{r-x}\left( \frac{A}{x} - \frac{B}{1-x} - \frac{C}{r-x} + xD - E \right) \right\} P_n^{(\mu,\nu)}(x)$$

where we have used the identities $\frac{b-x}{a-x} = 1 + \frac{b-a}{a-x}$ and $\frac{b-x}{(a-x)^2} = \frac{1}{a-x} + \frac{b-a}{(a-x)^2}$. The differential property of the Jacobi polynomial,

$$x(1-x)\frac{d}{dx} P_n^{(\mu,\nu)}(x) = (n+\mu+\nu+1)\left[ \frac{(\mu-\nu)n}{(2n+\mu+\nu)(2n+\mu+\nu+2)} P_n^{(\mu,\nu)}(x) \right.$$

$$\left. + \frac{(n+\mu)(n+\nu)}{(2n+\mu+\nu)(2n+\mu+\nu+1)} P_{n-1}^{(\mu,\nu)}(x) - \frac{n(n+1)}{(2n+\mu+\nu+1)(2n+\mu+\nu+2)} P_{n+1}^{(\mu,\nu)}(x) \right] \quad (16)$$

turns Eq. (15) into the following

$$\mathcal{J}\phi_n(x) = g_n x^\alpha (1-x)^\beta (r-x)^{\gamma+1} e^{-\delta x} \left\{ (n+\mu+\nu+1)\left( \frac{2\alpha + a - \nu - 1}{x} - \frac{2\beta + b - \mu - 1}{1-x} - \frac{2\gamma + c}{r-x} + d - 2\delta \right) \times \right.$$

$$\left[ \frac{(\mu-\nu)n}{(2n+\mu+\nu)(2n+\mu+\nu+2)} P_n^{(\mu,\nu)} + \frac{(n+\mu)(n+\nu)}{(2n+\mu+\nu)(2n+\mu+\nu+1)} P_{n-1}^{(\mu,\nu)} - \frac{n(n+1)}{(2n+\mu+\nu+1)(2n+\mu+\nu+2)} P_{n+1}^{(\mu,\nu)} \right]$$

$$-\left[ n(n+\mu+\nu+1) + (\alpha+\beta+\gamma)(\alpha+\beta+\gamma+a+b+c-1) + (2\delta-d)(\alpha+\gamma-r\gamma) + \delta(a+c-rc) + D \right] P_n^{(\mu,\nu)} \quad (17)$$

$$+\delta(\delta-d)x(1-x)P_n^{(\mu,\nu)} + x\left[ (2\delta-d)(\alpha+\beta+\gamma) + \delta(a+b+c) \right] P_n^{(\mu,\nu)} + \frac{1}{r-x}\left[ (d-2\delta)r(r-1)\gamma - \delta r(r-1)c \right.$$

$$\left. -\alpha(\alpha-1+a) + \beta(\beta-1+b) + (2r-1)\gamma(\gamma-1+c) + r(2\beta\gamma+\beta c+\gamma b) + (r-1)(2\alpha\gamma+\alpha c+\gamma a) + rD - E \right] P_n^{(\mu,\nu)}$$

$$\left. + \frac{1}{r-x}\left[ \frac{A + r\alpha(\alpha-1+a)}{x} - \frac{B - (r-1)\beta(\beta-1+b)}{1-x} - \frac{C + r(r-1)\gamma(\gamma-1+c)}{r-x} \right] P_n^{(\mu,\nu)} \right\}$$

where we have also used the identities $\frac{1}{x} = \frac{1}{r-x}\left( \frac{r}{x} - 1 \right)$ and $\frac{1}{1-x} = \frac{1}{r-x}\left( 1 + \frac{r-1}{1-x} \right)$. Now, the tridiagonal requirement of the TRA together with the recursion relation of the Jacobi polynomials allow only $x$-independent terms to multiply $P_{n\pm 1}^{(\mu,\nu)}(x)$ in (17) and only linear terms in $x$ multiplying $P_n^{(\mu,\nu)}(x)$. These constraints lead to the following basis parameters assignments

$$\mu^2 = (1-b)^2 + \frac{4B}{r-1}, \quad \nu^2 = (1-a)^2 - 4\frac{A}{r}, \quad 2\alpha = \nu + 1 - a, \quad (18a)$$

$$2\beta = \mu + 1 - b, \quad 2\gamma = -c, \quad \delta = \{0, d\}, \quad (18b)$$

and the following differential equation parameters constraints and relations

−5−

$$4A \le r(1-a)^2, \quad 4B \ge -(r-1)(1-b)^2, \quad 4C = r(r-1)c(c-2). \tag{19a}$$

$$E = \frac{A}{r} + \frac{B}{r-1} + rD - \frac{rc}{2}[a+b+c+d(r-1)-2] + \frac{c}{2}\left(a+\frac{c}{2}-1\right) \tag{19b}$$

These conditions reduce the number of independent parameters in the differential equation (6) from ten to eight $\{a,b,c,d,r,A,B,D\}$. Substitution of the parameters as given by Eq. (18) and Eq. (19) into Eq. (17) maps it into the following

$$\mathcal{J}\phi_n(x) = g_n x^\alpha (1-x)^\beta (r-x)^{\gamma+1} e^{-\delta x} \Bigg\{ \pm d(n+\mu+\nu+1) \times$$

$$\left[\frac{(\mu-\nu)n}{(2n+\mu+\nu)(2n+\mu+\nu+2)}P_n^{(\mu,\nu)} + \frac{(n+\mu)(n+\nu)}{(2n+\mu+\nu)(2n+\mu+\nu+1)}P_{n-1}^{(\mu,\nu)} - \frac{n(n+1)}{(2n+\mu+\nu+1)(2n+\mu+\nu+2)}P_{n+1}^{(\mu,\nu)}\right] \tag{20}$$

$$-\left[\left(n+\frac{\mu+\nu+1}{2}\right)^2 - \frac{1}{4}(a+b+c-1)^2 + \frac{d}{2}[a+c-rc \mp (\nu+1)] + D\right]P_n^{(\mu,\nu)} + \frac{d}{2}[a+b+c \mp (\mu+\nu+2)]xP_n^{(\mu,\nu)}\Bigg\}$$

where the top and bottom signs correspond to $\delta = 0$ and $\delta = d$, respectively. Thus, the exponential factor $e^{-\delta x}$ in the basis (8) becomes equal to either 1 or $e^{-xd}$, respectively. Next, we use the three-term recursion relation of the Jacobi polynomials,

$$x P_n^{(\mu,\nu)}(x) = \frac{1}{2}\left[\frac{\nu^2-\mu^2}{(2n+\mu+\nu)(2n+\mu+\nu+2)}+1\right]P_n^{(\mu,\nu)}(x)$$

$$+ \frac{(n+\mu)(n+\nu)}{(2n+\mu+\nu)(2n+\mu+\nu+1)}P_{n-1}^{(\mu,\nu)}(x) + \frac{(n+1)(n+\mu+\nu+1)}{(2n+\mu+\nu+1)(2n+\mu+\nu+2)}P_{n+1}^{(\mu,\nu)}(x) \tag{21}$$

in Eq. (20) to obtain

$$\frac{2}{d}\mathcal{J}\phi_n(x) = g_n x^\alpha (1-x)^\beta (r-x)^{\gamma+1} e^{-\delta x} \Bigg( \Bigg\{ -\frac{2}{d}\left[\left(n+\frac{\mu+\nu+1}{2}\right)^2 - \frac{1}{4}(a+b+c-1)^2 + D\right] - a + (r-1)c \pm (\nu+1)$$

$$\pm \frac{2(\mu-\nu)n(n+\mu+\nu+1)}{(2n+\mu+\nu)(2n+\mu+\nu+2)} + \frac{1}{2}\left[1+\frac{\nu^2-\mu^2}{(2n+\mu+\nu)(2n+\mu+\nu+2)}\right][a+b+c \mp (\mu+\nu+2)]\Bigg\} P_n^{(\mu,\nu)}(x)$$

$$+ \frac{(n+\mu)(n+\nu)}{(2n+\mu+\nu)(2n+\mu+\nu+1)}[a+b+c \pm (2n+\mu+\nu)]P_{n-1}^{(\mu,\nu)}(x) \tag{22}$$

$$+ \frac{(n+1)(n+\mu+\nu+1)}{(2n+\mu+\nu+1)(2n+\mu+\nu+2)}[a+b+c \mp (2n+\mu+\nu+2)]P_{n+1}^{(\mu,\nu)}(x) \Bigg)$$

If we choose the normalization constant as $g_n = (2n+\mu+\nu+1)\frac{\Gamma(n+\mu+\nu+1)}{\Gamma(n+\nu+1)}$ and rewrite this equation in terms of the basis elements $\{\phi_n(x)\}$ instead of $\{P_n^{(\mu,\nu)}(x)\}$ (i.e., taking care of the normalization factor $g_n$) then we obtain

$$\frac{\mp 1/d}{r-x}\mathcal{J}\phi_n(x)=\left\{\pm\frac{1}{d}\left[\left(n+\frac{\mu+\nu+1}{2}\right)^2-\left(\chi-\frac{1}{2}\right)^2+D\right]-\left(\frac{\nu+1}{2}\mp\chi\right)\mp\frac{b+rc}{2}\right.$$
$$+\frac{(\nu-\mu)n(n+\mu+\nu+1)}{(2n+\mu+\nu)(2n+\mu+\nu+2)}+\frac{1}{2}\left[1+\frac{\nu^2-\mu^2}{(2n+\mu+\nu)(2n+\mu+\nu+2)}\right]\left(\frac{\mu+\nu}{2}+1\mp\chi\right)\right\}\phi_n(x) \quad (23)$$
$$-\frac{(n+\mu)(n+\mu+\nu)}{(2n+\mu+\nu)(2n+\mu+\nu-1)}\left(n+\frac{\mu+\nu}{2}\pm\chi\right)\phi_{n-1}(x)+\frac{(n+1)(n+\nu+1)}{(2n+\mu+\nu+3)(2n+\mu+\nu+2)}\left(n+\frac{\mu+\nu}{2}+1\mp\chi\right)\phi_{n+1}(x)$$

where $\chi:=\frac{1}{2}(a+b+c)$. Consequently, the differential equation $\mathcal{J}\hat{y}(x)=0$ becomes the following recursion relation for the expansion coefficients $\{f_n(z)\}$

$$\frac{1}{2}\left[\mu+1\pm(b+rc)\right]f_n(z)=\left\{\pm\frac{1}{d}\left[\left(n+\frac{\mu+\nu+1}{2}\right)^2-\left(\chi-\frac{1}{2}\right)^2+D\right]+\mu+1\right.$$
$$+\frac{(\nu-\mu)n(n+\mu+\nu+1)}{(2n+\mu+\nu)(2n+\mu+\nu+2)}-\frac{1}{2}\left[1+\frac{\mu^2-\nu^2}{(2n+\mu+\nu)(2n+\mu+\nu+2)}\right]\left(\frac{\mu+\nu}{2}+1\mp\chi\right)\right\}f_n(z) \quad (24)$$
$$-\frac{(n+\mu+1)(n+\mu+\nu+1)}{(2n+\mu+\nu+1)(2n+\mu+\nu+2)}\left(n+\frac{\mu+\nu}{2}+1\pm\chi\right)f_{n+1}(z)$$
$$+\frac{n(n+\nu)}{(2n+\mu+\nu)(2n+\mu+\nu+1)}\left(n+\frac{\mu+\nu}{2}\mp\chi\right)f_{n-1}(z)$$

As noted above, writing $f_n(z)=f_0(z)P_n(z)$ makes $P_0(z)=1$ and this recursion relation gives $P_n(z)$ as a polynomial of degree $n$ in some proper argument $z$ with a positive definite weight function as $[f_0(z)]^2$ [2,18]. The recursion coefficients that multiply $f_{n\pm 1}(z)$ are ratios of cubic to quadratic polynomials in $n$. The same property is found in the recursion relation of the continuous Hahn polynomial, $\tilde{p}_n(z):=\tilde{p}_n(z;\lambda,\tau,\rho,\sigma)={}_3F_2\left({-n,n+\lambda+\tau+\rho+\sigma-1,\lambda+iz \atop \lambda+\rho,\lambda+\sigma}\bigg|1\right)$ [19], that reads

$$(\lambda+iz)\tilde{p}_n(z)=\mathcal{A}_n\tilde{p}_{n+1}(z)+\mathcal{C}_n\tilde{p}_{n-1}(z)-(\mathcal{A}_n+\mathcal{C}_n)\tilde{p}_n(z), \quad (25)$$

where

$$\mathcal{A}_n=-\frac{(n+\lambda+\tau+\rho+\sigma-1)(n+\lambda+\rho)(n+\lambda+\sigma)}{(2n+\lambda+\tau+\rho+\sigma-1)(2n+\lambda+\tau+\rho+\sigma)}, \quad (26a)$$

$$\mathcal{C}_n=\frac{n(n+\tau+\rho-1)(n+\tau+\sigma-1)}{(2n+\lambda+\tau+\rho+\sigma-2)(2n+\lambda+\tau+\rho+\sigma-1)}. \quad (26b)$$

Comparing (24) to (25) shows that the sought after polynomial $P_n(z)$ is a modified version of the continuous Hahn polynomial obtained by deforming the recursion relation (25) as follows

$$(\lambda+iz)\tilde{p}_n^d(z)=\mathcal{A}_n\tilde{p}_{n+1}^d(z)+\mathcal{C}_n\tilde{p}_{n-1}^d(z)-(\mathcal{A}_n+\mathcal{C}_n\mp d^{-1}\mathcal{B}_n)\tilde{p}_n^d(z), \quad (27)$$

where $\mathcal{B}_n = \left(n + \frac{\mu+\nu+1}{2}\right)^2 - \left(\chi - \frac{1}{2}\right)^2 + D$ and the deformation parameter is $\pm d^{-1}$. Moreover, the modified polynomial parameters are given in terms of the differential equation parameters as follows

$$\lambda + \mathrm{i} z = \frac{1}{2}\left[\mu + 1 \pm (b + rc)\right], \qquad \sigma = -\lambda + \mu + 1. \tag{28a}$$

$$\tau = \lambda + \frac{\nu - \mu}{2} \mp \chi, \qquad \rho = -\lambda + \frac{\mu + \nu}{2} + 1 \pm \chi. \tag{28b}$$

Note that the deformation parameter $d$ appears in the recursion relation only as shown explicitly in (27) and nowhere else in the relation. Therefore, the original recursion relation of the continuous Hahn polynomial is recovered in the limit $d \to \infty$. Finally, the solution of the ten-parameter 2nd order linear differential equation (6) becomes the following series

$$y(x+\alpha) = x^\alpha (1-x)^\beta (r-x)^\gamma e^{-\delta x} \sqrt{\omega^d(z)}$$
$$\sum_{n=0}^{\infty} (2n + \mu + \nu + 1) \frac{\Gamma(n+\mu+\nu+1)}{\Gamma(n+\nu+1)} \tilde{p}_n^d(z; \lambda, \tau, \rho, \sigma) P_n^{(\mu,\nu)}(x) \tag{29}$$

where the parameters $\{\alpha, \beta, \gamma, \delta, \mu, \nu\}$ and $\{\lambda, \tau, \rho, \sigma\}$ are given in terms of the differential equation parameters as shown by Eq. (18) and Eq. (28), respectively. The positive definite function $\omega^d(z)$, which is identical to $[f_0(z)]^2$, is the weight function of the modified continuous Hahn polynomial $\tilde{p}_n^d(z)$. This weight function, however, is yet to be derived analytically by experts in the field of orthogonal polynomials. Assuming that this could only be done under the same positivity conditions as the original continuous Hahn polynomial [19], then we require that $\rho = \lambda$ and $\sigma = \tau$ (i.e., $\mu = \nu$ and $2\lambda = \mu + 1 \pm \chi$). For this special case, the differential equation parameters must meet the following restriction in addition to those given by Eq. (19)

$$\frac{A}{r} + \frac{B}{r-1} = \frac{a-b}{2}\left(\frac{a+b}{2} - 1\right). \tag{30}$$

This reduces further the number of independent parameters of the differential equation to seven.

## III. Conclusion

In this work, we introduced a ten-parameter ordinary linear differential equation of the second order with four singular pints. Three of these are finite and regular whereas one of them is irregular at infinity. We proposed a solution as an infinite series of square integrable functions written in terms of the Jacobi polynomials. Using the TRA, we showed that if two of the equation parameters ($C$ and $E$) are constrained by a specific relation to the rest as shown in Eq. (19) then the set of expansion coefficients of the series becomes a modified version of the continuous Hahn polynomial. We identified the corresponding three-term recursion relation that will give all of the modified polynomials $\{\tilde{p}_n^d(z)\}$ to any degree starting with the initial value $\tilde{p}_0^d(z) = 1$.

Due to the anticipated significance of the polynomial $\tilde{p}_n^d(z)$ to the solution of various problems in physics and mathematics that are associated with the differential equation (3), we call upon experts in the field of orthogonal polynomials to derive their properties (weight function, generating function, orthogonality, zeros, asymptotics, Rodrigues-type formula, differential or shift formula, etc.).